\documentclass[amsmath,amssymb,pra,aps,showpacs,superscriptaddress,twocolumn]{revtex4-1}
\usepackage{amsmath,amsfonts,amssymb,amsthm,graphics,graphicx,epsfig,bbm}
\usepackage[colorlinks=true,citecolor=blue,linkcolor=blue,urlcolor=blue]{hyperref}
\usepackage[usenames]{color}
\usepackage{graphicx}
\usepackage{subfigure}
\usepackage{amsmath}
\usepackage{epsfig}
\usepackage{dcolumn}
\usepackage{bm}
\usepackage{color}
\usepackage{times}
\usepackage{epstopdf}
\usepackage{amssymb}
\usepackage{amstext}
\usepackage{latexsym}
\usepackage{hyperref}
\usepackage{amsfonts}
\usepackage{psfrag}
\usepackage{soul,xcolor}
\usepackage[normalem]{ulem}
\usepackage{dsfont}
\usepackage{adjustbox}
\usepackage{txfonts}

\newcommand{\ket}[1]{\vert #1 \rangle}
\newcommand{\bra}[1]{\langle #1 \vert}
\newcommand{\ketbra}[2]{\vert #1 \rangle \langle #2 \vert}

\begin{document}

\title{Parity-assisted generation of nonclassical states of light in circuit quantum electrodynamics}

\author{F. A. C{\'a}rdenas-L{\'o}pez}
\email[Corresponding authors:~~]{\\francisco.cardenas@usach.cl.\\}
\affiliation{Departamento de F\'isica, Universidad de Santiago de Chile (USACH), Avenida Ecuador 3493, 9170124, Santiago, Chile}
\affiliation{Center for the Development of Nanoscience and Nanotechnology 9170124, Estaci\'on Central, Santiago, Chile}
	
\author{G. Romero}
\affiliation{Departamento de F\'isica, Universidad de Santiago de Chile (USACH), Avenida Ecuador 3493, 9170124, Santiago, Chile}

\author{L. Lamata}
\affiliation{Department of Physical Chemistry, University of the Basque Country UPV/EHU, Apartado 644, 48080 Bilbao, Spain}

\author{E. Solano}
\affiliation{Department of Physical Chemistry, University of the Basque Country UPV/EHU, Apartado 644, 48080 Bilbao, Spain}
\affiliation{IKERBASQUE, Basque Foundation for Science, Maria Diaz de Haro 3, 48013 Bilbao, Spain}
\affiliation{Department of Physics, Shanghai University, 200444 Shanghai, China}

\author{J. C. Retamal}
\affiliation{Departamento de F\'isica, Universidad de Santiago de Chile (USACH), Avenida Ecuador 3493, 9170124, Santiago, Chile}
\affiliation{Center for the Development of Nanoscience and Nanotechnology 9170124, Estaci\'on Central, Santiago, Chile}

\pacs{Microwave photons, quantum entanglement, superconducting circuits, circuit quantum electrodynamics, quantum Rabi model.}
\date{\today}

\begin{abstract}
We propose a method to generate nonclassical states of light in multimode microwave cavities. Our approach considers two-photon processes that take place in a system composed of two extended cavities and an ultrastrongly coupled light-matter system. Under specific resonance conditions, our method generates, in a deterministic manner, product states of uncorrelated photon pairs, Bell states, and W states. We demonstrate improved generation times when increasing the number of multimode cavities, and prove the generation of genuine multipartite entangled states when coupling an ancillary system to each cavity. Finally, we discuss the feasibility of our proposal in circuit quantum electrodynamics.   
\end{abstract}
\maketitle
\section{Introduction}\label{Introduction}
The state-of-the-art of devices exhibiting quantum behaviour has grown extensively in the last two decades. Remarkable platforms such as superconducting circuits \cite{Nature.474.589,Nature.Phys.8.292,Science.339.1169} and circuit quantum electrodynamics (QED) \cite{Nature.431.162,Phys.Rev.A.69.062320} have allowed the implementation of microwave quantum photonics~\cite{Phot.544,Phys.Rep}, where superconducting electrical circuits mimic the behavior of atoms and cavities \cite{Phys.Rev.B.60.15398,Phys.Rev.A.76.042319,Jour.Appl.Phys.104.113904}. 
In this manner, the capability of tailoring internal circuit parameters to obtain devices with long coherence times and switchable coupling strengths yielded quantum optics experiments such as electromagnetically induced transparency \cite{Phys.Rev.Lett.104.193601}, photon blockade \cite{Phys.Rev.Lett.106.243601}, and lately to manipulate the parity symmetric of an artificial atom in situ \cite{Phys.Rev.Lett.121.060503} to name a few.  A distinctive aspect of microwave photonics is the inherent nonlinearity coming from Josephson junction devices that makes possible to build photonic crystals with Kerr and Cross-Kerr nonlinearities much larger than the one observed in optical devices
\cite{Nat.Phys.6.296,Phys.Rev.Lett.105.100504,Phys.Rev.A.86.013814,New.Jour.Phys.14.075024}. 
This allows for enhancing processes such as parametric down conversion \cite{Phis.Rev.B.76.205416,Phys.Rev.A.79.013804,Sci.Rep.4.7289,Phys.Rev.A.94.053814}, and the generation of nonclassical states of light \cite{Phys.Rev.A.39.2519,Phys.Rev.A.69.043804,Phys.Rev.Lett.101.253602,Phys.Rev.Lett.95.140504,Phys.Rev.A.89.013820}. Likewise, the notable features of superconducting circuits have also triggered a bunch of proposals for microwave photon generation in systems composed of a large number of cavities. In this context, it is possible to find proposals for the generation of entangled photon states such as NOON and MOON \cite{Phys.Rev.Lett.109.210501,Sci.Rep.6.236.46,New.Jour.Phys.12.093036,Phys.Rev.Lett.105.050501,Phys.Rev.Lett.106.060401} states, studies  of correlated photons emitted from a cascade system \cite{Phys.Rev.Lett.119.140504}, as well as the implementation of a CNOT gate between qubits encoded in a cavity \cite{arXiv:1712.05854}, among other applications \cite{arXiv:1709.05425,Phys.Rev.X.6.031036,arXiv:1712.08593}. 

On the other hand, circuit QED has also made possible to achieve light-matter coupling strengths such as the ultrastrong (USC) \cite{PhysRevA.80.032109,Nature.6.772,Phys.Rev.Lett.105.237001,New.Jour.Phys.19.023022,Nat.Phys.13.39} and deep-strong (DSC)~\cite{PhysRevLett.105.263603,Nature.3906} regimes of light-matter coupling~\cite{USCreview}. In both cases, 
as the coupling strength between the light and matter becomes comparable (USC) or larger than the frequency of the field mode (DSC), the rotating wave approximation breaks down and the simplest model that describes the physical situation is the quantum Rabi model~\cite{Phys.Rev.49.624, Phys.Rev.Lett.107.100401,USCreview}. This model exhibits a discrete parity symmetry and an anharmonic energy spectrum that provide a set of resources for quantum information tasks and quantum simulations \cite{PhysRevLett.107.190402,PhysRevLett.108.120501,srep08621,srep11818,PhysRevB.91.064503,PhysRevA.94.012328,Phys.Rev.A.97.022306}.

Based on the latest developments in superconducting circuits, here, we propose a method to generate nonclassical states of light in multimode microwave cavities. Our approach considers two-photon 
processes taking place in a system composed of two extended cavities and an ultrastrongly coupled light-matter system, hereafter called quantum Rabi system (QRS). Under specific resonance conditions, our method allows a deterministic generation of identical photonic quantum states of different frequency which can be uncorrelated photon state or correlated Bell and W states. Furthermore, we could extend our protocol to more cavities. In this sense, the generation time of these nonclassical states is inversely proportional to the number of cavities in the system. This collective effect is due to the multimode configuration of our setup. On the other hand, we show the generation of genuine multipartite entangled states when coupling an ancillary system to each cavity. Finally, we propose a physical implementation of our scheme considering near-term technology of superconducting circuits.

This paper is organized as follows: In section~\ref{The model}, we introduce our physical scheme. 
In section~\ref{Quantum Rabi model energy spectrum and the parity symmetry}, we discuss about the main aspects of the physics of QRS, that is, its parity symmetry and the underlying selection rules for state transitions. In section~\ref{Multiphoton generation mediated by the quantum Rabi model}, we discuss the two-photon processes presented in our physical system, and the generation of nonclassical states of light. In section \ref{copies}, we show that our model allows for generating copies of density matrices. In section \ref{swapping}, we study swapping processes for the generation of genuine multipartite entanglement. In section \ref{implementation}, we present a physical implementation of our method in superconducting circuits. Finally, in section~\ref{conclusion}, we present our concluding remarks.

\section{The Model}\label{The model}
Let us consider a two-level system of frequency $\omega_{q}$ interacting with a quantized 
electromagnetic field mode of frequency $\omega_{\rm{cav}}$ in the USC regime. This system is described by the quantum Rabi Hamiltonian \cite{Phys.Rev.49.624,Phys.Rev.Lett.107.100401} ($\hbar=1$)
\begin{eqnarray}
\label{Eq1}
\mathcal{H}_{\rm{QRS}} = \omega_{\rm{cav}}a^{\dag}a + \frac{\omega_{q}}{2}\sigma^{z} +  g\sigma^{x}(a^{\dag} + a). 
\end{eqnarray}
Here, $a^{\dag}(a)$ is the creation (annihilation) boson operator for the field mode, the operators $\sigma^{x}$ and $\sigma^{z}$ are the Pauli matrices describing the two-level system, and $g$ is the light-matter coupling strength. In addition, $N$ multimode resonators \cite{multimode2}, each supporting $M=2$ modes of frequencies $\omega_{1}^{\ell}$ and $\omega_{2}^{\ell}$, are coupled to the edges of the QRS through field quadratures. Notice that each mode couples to the QRS with coupling strengths $J_{1}^{\ell}$ and $J_{2}^{\ell}$, respectively. This physical situation will be described by the Hamiltonian 
\begin{eqnarray}
\label{Eq2}
\mathcal{H} &=& \mathcal{H}_{\rm{QRS}} + \mathcal{H}_{\rm{c}} + \mathcal{H}_{I},\\
\label{Eq3}
\mathcal{H}_{\rm{c}} &=& \sum_{\ell=1}^{N}(\omega_{1}^{\ell}b^{\dag}_{\ell}b_{\ell} + \omega_{2}^{\ell}c^{\dag}_{\ell}c_{\ell}),\\
\label{Eq4}
\mathcal{H}_{I}&=& \sum_{\ell=1}^{N} \big[J_{1}^{\ell}(b_{\ell}^{\dag}+b_{\ell}) + J_{2}^{\ell}(c_{\ell}^{\dag}+c_{\ell})\big](a + a^{\dag}),
\end{eqnarray}
where $b_{\ell}^{\dag}(b_{\ell})$ and $c_{\ell}^{\dag}(c_{\ell})$ are the creation (annihilation) boson operators 
for the first and second field mode of the $\ell$th cavity, respectively. Notice that the coupling strength between 
resonators $J^{\ell}_{1,2}$ can be several orders of magnitude smaller than $\omega_{1,2}^{\ell}$ \cite{Underwood2012}. 
Hence, the counter-rotating terms present in Eq.~(\ref{Eq4}) can be neglected through the rotating wave approximation (RWA) leading to the 
following interaction Hamiltonian 
\begin{eqnarray}
\label{Eq5}
\mathcal{H}_{I}=\sum_{\ell=1}^{N} \big[(J_{1}^{\ell}b_{\ell} + J_{2}^{\ell}c_{\ell})a^{\dag} + (J_{1}^{\ell}b_{\ell}^{\dag} + J_{2}^{\ell}c_{\ell}^{\dag})a\big].
\end{eqnarray}
In what follows, we will discuss the features of the energy spectrum of the QRS, that is, its anharmonicity and the internal symmetry arising in the USC regime.
\section{Parity symmetry $\mathbb{Z}_{2}$ and selection rules}\label{Quantum Rabi model energy spectrum and the parity symmetry}
\begin{figure}[t!]
\centering
\includegraphics[width=1\linewidth]{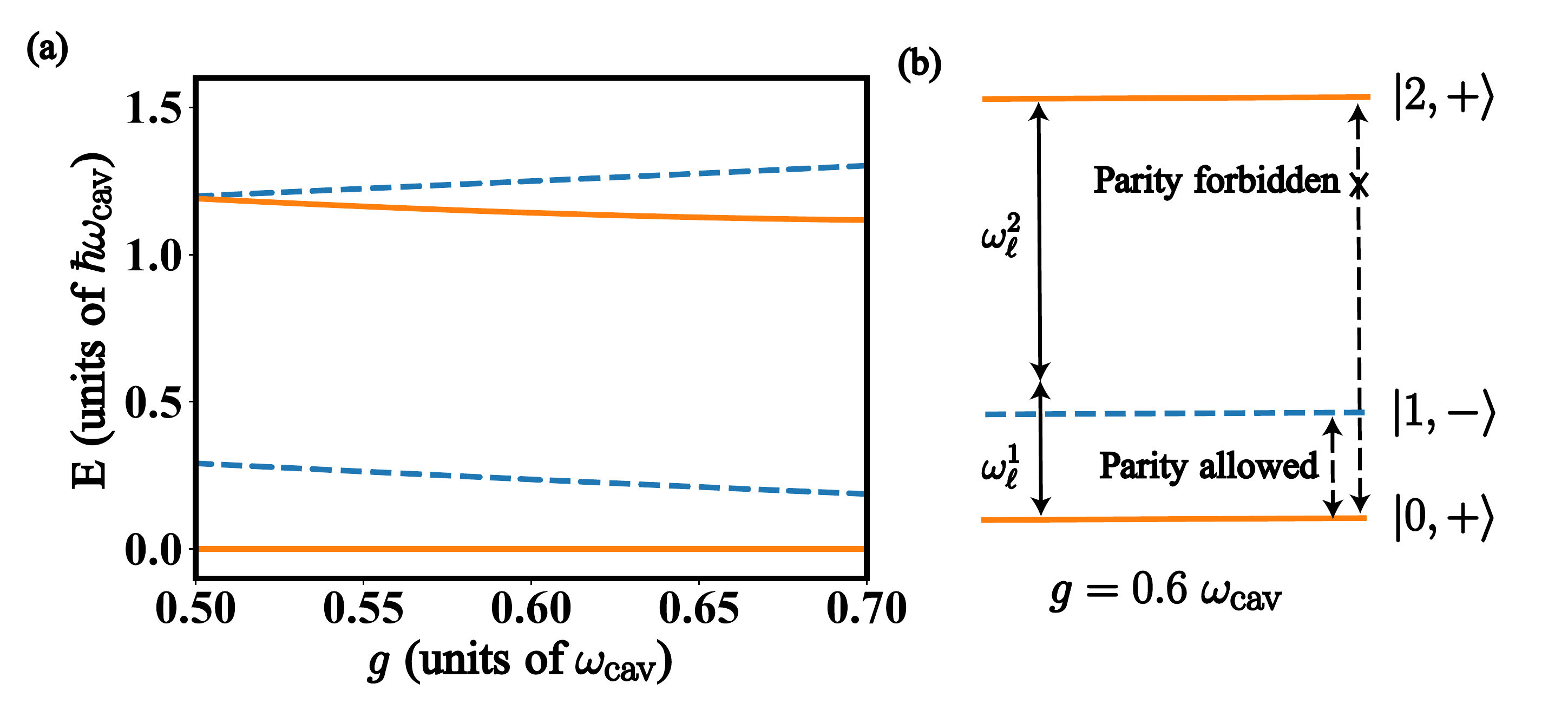}
\caption{(Color online) (a) Energy spectrum of the Hamiltonian in Eq.~(\ref{Eq1}) as a function of the coupling strength $g$. 
Blue dashed lines stand for states with parity $p=+1$. Orange continuous lines correspond to states with parity $p=-1$. 
(b) Diagram of the energy levels at $g=0.6~\omega_{\rm{cav}}$. In these numerical calculations
we use $\omega_{q}=0.8~\omega_{\rm{cav}}$.}
\label{fig:fig1}
\end{figure}
The energy spectrum of the QRS presents interesting features which have proven useful in performing 
quantum information processing
 \cite{PhysRevLett.107.190402,PhysRevLett.108.120501,srep08621,srep11818,PhysRevB.91.064503,PhysRevA.94.012328}. These features correspond to the anharmonicity of the energy levels and the selection rules imposed by the $\mathbb{Z}_2$ symmetry arising in the USC regime. In Fig. \ref{fig:fig1}, we show the first four energy levels of the QRS as a function of $g/\omega_{\rm cav}$, where we see an anharmonic energy spectrum. Moreover, in the QRS, it is possible to define the parity operator $\mathcal{P}=-\sigma^{z}\otimes e^{i\pi a^{\dag}a}$ which has discrete spectrum $p=\pm1$. Notice that $\mathcal{P}$ commutes with the QRS Hamiltonian, $[H_{\rm QRS},\mathcal{P}]=0$, thus enabling the diagonalization of both operators in a common basis $\{\ket{\sigma,p}\}_{\sigma=0}^{\infty}$. We label each quantum state regarding two quantum numbers, $\sigma$ corresponds to the energy level while $p$ denotes its parity value. In Fig. \ref{fig:fig1}, states with parity $+1$($-1$) are denoted by the continuous orange (dashed blue) line. As a consequence, the Hilbert space of the QRS is divided into two parts, the even and the odd parity subspaces. This allows, depending on the kind of driving, the possibility of connecting states with different or equal parity. For instance, it has been proven that drivings like $\mathcal{H}_{D}\sim(a^{\dag}+a)$ and $\mathcal{H}_{D}\sim\sigma^{x}$ connect states belonging to different subspaces \cite{PhysRevA.94.012328}. This happens because the matrix element $\bra{\sigma,\pm}\mathcal{H}_{D}\ket{\sigma',\mp}\neq0$. Moreover, for a driving like $\mathcal{H}_{D}\sim\sigma^{z}$, only states with equal parity can be connected since the matrix
element $\bra{\sigma,\pm}\mathcal{H}_{D}\ket{\sigma',\pm}\neq0$. 

\section{Two photon process mediated by the quantum Rabi system}\label{Multiphoton generation mediated by the quantum Rabi model}
\begin{figure}[t!]
\centering
\includegraphics[width=1\linewidth]{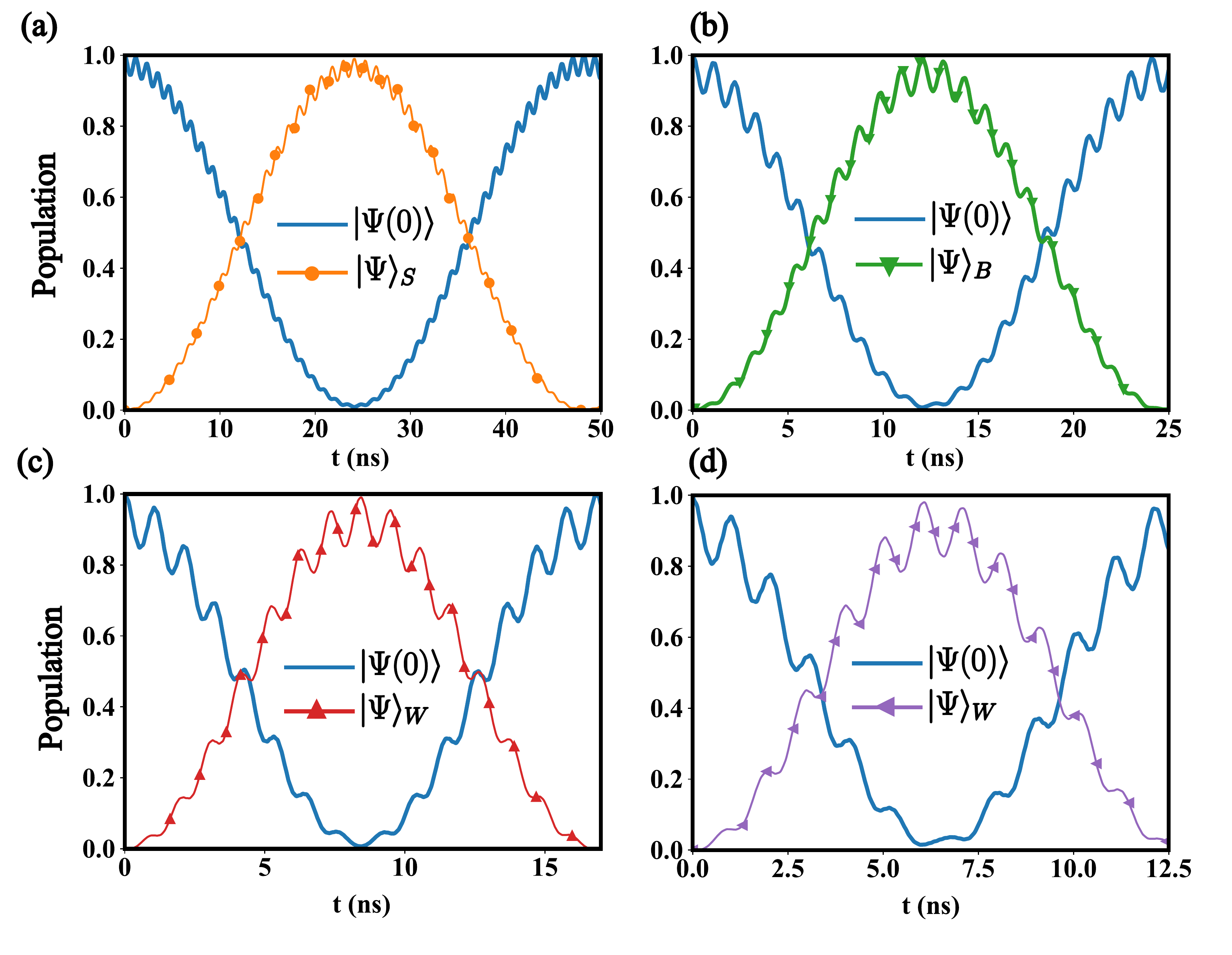}
\caption{(Color online). Population evolution of the \textit{ab initio} model Eq. (\ref{Eq2}) for initial state $\ket{\Psi(0)}=\ket{2,+}\bigotimes_{\ell,n}^{N,M}\ket{0_{\ell}^{n}}$ with cases $N=1$ (a), $N=2$ (b), $N=3$ (c), and $N=4$ (d) multimode cavities. Blue continuous line is the evolution of the initial state $\ket{\Psi(0)}$. (a) Orange dotted line denotes the population of $\ket{\Psi}_S=\ket{0,+}\otimes\ket{1_{\omega_{1}}}\otimes\ket{1_{\omega_{2}}}$. (b)  Green dotted line stands for the population of $\ket{\Psi}_B=\ket{0,+}\otimes\ket{\Psi^{+}_{\omega_{1}}}\otimes\ket{\Psi^{+}_{\omega_{2}}}$, and (c) red dotted line stands for $\ket{\Psi}_W=\ket{0,+}\otimes\ket{W_{\omega_{1}}}\otimes\ket{W_{\omega_{2}}}$. The parameters for these calculations can be found in the main text.}
\label{fig:fig2}
\end{figure}
Here, we propose the implementation of a two-photon process mediated by the QRS, which relies on its anharmonicity and the selection rules previously discussed. In particular, we provide specific resonance conditions between multimode cavities and the QRS to achieve the phase matching condition analogue to the usual parametric down-conversion process in optical systems.

Let us consider the following set of parameters for the QRS $\omega_{q}=0.8~\omega_{\rm{cav}}$ and $g=0.6~\omega_{\rm{cav}}$. In this case, as shown in Fig \ref{fig:fig1}, the first three energy levels form a cascade $\Xi$ system similar to Rydberg atoms studied in cavity quantum electrodynamics \cite{Phys.Rev.A 35.154}. The ground and second excited state have parity $p=+1$, while the first excited state has parity $p=-1$, see Fig \ref{fig:fig1}(b). According to the type of interaction of the multimode cavities with the QRS, see Eq. (\ref{Eq4}), a single photon will not be able to produce a transition between the second excited state $\ket{2,+}$ and the ground state $\ket{0,+}$ since it is forbidden by parity. However, these states can be connected through a second-order process. The latter may occur when the sum of frequencies of the modes, belonging to a cavity, matches that of the energy transition between the ground and the second excited state of the QRS, i.e. $\omega_{1}^{\ell}+\omega_{2}^{\ell}=\nu_{20}$. Moreover, the frequency of each mode must be far-off-resonance with respect to the frequency of the first excited state $\omega_{1,2}^{\ell}\gg\nu_{10}$. Under these conditions, the intermediate level can be adiabatically eliminated leading to the effective Hamiltonian
\begin{eqnarray}
\label{Eq6}
\mathcal{H}_{\rm{eff}}^{\ell}=\mathcal{H}_{\rm{QRS}} + \mathcal{H}_{\rm{c}}+\sum_{\ell,\ell'=1}^{N}\mathcal{J}_{\ell}^{\ell'}(b_{\ell}^{\dag}c_{\ell'}^{\dag}\mathcal{S}^{-} + b_{\ell}c_{\ell'}\mathcal{S}^{+}),
\end{eqnarray}
which describes simultaneous two-photon processes in both cavities. Here, $S^{+}=\ketbra{2,+}{0,+}$ corresponds to the ladder operator of the QRS in the effective two-level basis. Furthermore, the effective coupling strength $\mathcal{J}_{\ell}^{\ell'}$ is defined as follows
\begin{eqnarray}
\label{Eq10}
\mathcal{J}_{\ell}^{\ell'} =J_{1}^{\ell}J_{2}^{\ell'}\chi_{01}\chi_{21}\bigg[\frac{1}{\Delta_{10}^{1}} + \frac{1}{\Delta_{21}^{2}}\bigg].
\end{eqnarray} 
Here, we define the matrix element of the operator $a$ in the QRS basis as $\chi_{kj}^{\pm}=\bra{k,+}a\ket{j,-}$ and the QRS-mode detuning $\Delta_{kj}^{1,2} =\omega_{1,2}^{\ell}-\nu_{kj}$. The Hamiltonian in Eq. (\ref{Eq6}) gives rise to several parametric down conversion processes mediated by the QRS, i.e., by starting with one excitation on the QRS of energy $\nu_{20}$, it may produce a pair of photons of frequencies $\omega_{1}$ and $\omega_{2}$. The photons generated by this scheme 
will distribute on the multimode cavities according to the relation $\omega_{1}^{\ell}+\omega_{2}^{\ell'}=\nu_{20}$. Depending on the number of cavities $N$, this condition enables us to generate two uncorrelated single-photons ($N=1$), or producing identical entangled states of different frequency such as Bell states ($N=2$) or $W$ states ($N\ge3$). For the cases, $N=\{1,2,3\}$ the effective Hamiltonians read
\begin{widetext}
\begin{subequations}
\label{Eq8}
\begin{eqnarray}
\mathcal{H}_{\rm{eff}}^{1}&=&\mathcal{J}_{2}^{1}\big[b^{\dag}_{1}c^{\dag}_{1}\mathcal{S}^{-}  + b_{1}c_{2}\mathcal{S}^{+}\big].\\ \mathcal{H}_{\rm{eff}}^{2}&=&\mathcal{J}_{2}^{1}\big[b^{\dag}_{1}c^{\dag}_{1} + b^{\dag}_{2}c^{\dag}_{2} + b^{\dag}_{1}c^{\dag}_{2} + b^{\dag}_{2}c^{\dag}_{1}\big]\mathcal{S}^{-}+\rm{H.c}.\\ 	\mathcal{H}_{\rm{eff}}^{3}&=&\mathcal{J}_{2}^{1}\big[b^{\dag}_{1}c^{\dag}_{1} + b^{\dag}_{2}c^{\dag}_{2} + b^{\dag}_{3}c^{\dag}_{3} + b^{\dag}_{1}c^{\dag}_{2}+b^{\dag}_{1}c^{\dag}_{3} + b^{\dag}_{2}c^{\dag}_{1} + b^{\dag}_{2}c^{\dag}_{3} + b^{\dag}_{3}c^{\dag}_{1} + b^{\dag}_{3}c^{\dag}_{2}\big]\mathcal{S}^{-} + \rm{H.c}.
\end{eqnarray}
\end{subequations}
\end{widetext}
The protocol works as follows: we initially consider the entire system in its ground state i.e., $\ket{\Psi(0)}=\ket{0,+}\bigotimes_{\ell,\ell'}^{N}\ket{0_{\ell},0_{\ell'}}$. Afterwards, one may excite the QRS with a microwave pulse with frequency $\nu=\nu_{20}$. This interaction can be modeled by the Hamiltonian $\mathcal{H}_{D}=\Omega\cos(\nu_{20}t)\sigma^{z}$. Notice that $\mathcal{H}_{D}$
preserves the $\mathbb{Z}_2$ symmetry of the QRS, thus enabling transitions between states of equal parity. The state of the system, after an interaction time $t=\pi/\Omega$, is given by $\ket{\Psi(\pi/\Omega)}=\ket{2,+}\bigotimes_{\ell,\ell'}^{N}\ket{0_{\ell},0_{\ell'}}$. Then, the system evolves under the Hamiltonian (\ref{Eq2}) for a time $t_S=\pi/(2\mathcal{J}_{2}^{1})$, $t_B=\pi/(4\mathcal{J}_{2}^{1})$, or $t_W=\pi/(6\mathcal{J}_{2}^{1})$, for generating uncorrelated single photons, pair of Bell states, or pair of W states, respectively. As a result, the QRS excitation generates two photons distributed on the cavities satisfying the relation $\omega_{1}^{\ell}+\omega_{2}^{\ell'}=\nu_{20}$. The wavefunctions of the system after algebraic manipulation read
\begin{subequations}
\label{Eq9}
\begin{eqnarray}
\ket{\Psi(\pi/\Omega + \pi/2\mathcal{J}_{2}^{1})}_S &=&\ket{+,0}\otimes\ket{1_{\omega_1}}\otimes\ket{1_{\omega_2}},\\
\ket{\Psi(\pi/\Omega + \pi/4\mathcal{J}_{2}^{1})}_B &=&\ket{+,0}\otimes\ket{\Psi^{+}_{\omega_1}}\otimes\ket{\Psi^{+}_{\omega_2}},\\
\ket{\Psi(\pi/\Omega + \pi/6\mathcal{J}_{2}^{1})}_W &=&\ket{+,0}\otimes\ket{W_{\omega_1}}\otimes\ket{W_{\omega_2}},
\end{eqnarray}
\end{subequations}
where $\ket{\Psi^{+}_{\omega_n}}$ is the Bell state for photons of frequency $\omega_n$ distributed over different resonators, that is, $\ket{\Psi^{+}_{\omega_n}}=\frac{1}{\sqrt{2}}[\ket{1_{\omega_n}}\ket{0_{\omega_n}}+\ket{0_{\omega_n}}\ket{1_{\omega_n}}]$. Also, the state $\ket{W_{\omega_n}}$ stands for a $W$ state of a single photon of frequency $\omega_n$ distributed over different cavities.  
\begin{figure*}[t]
\centering
\includegraphics[width=1\linewidth]{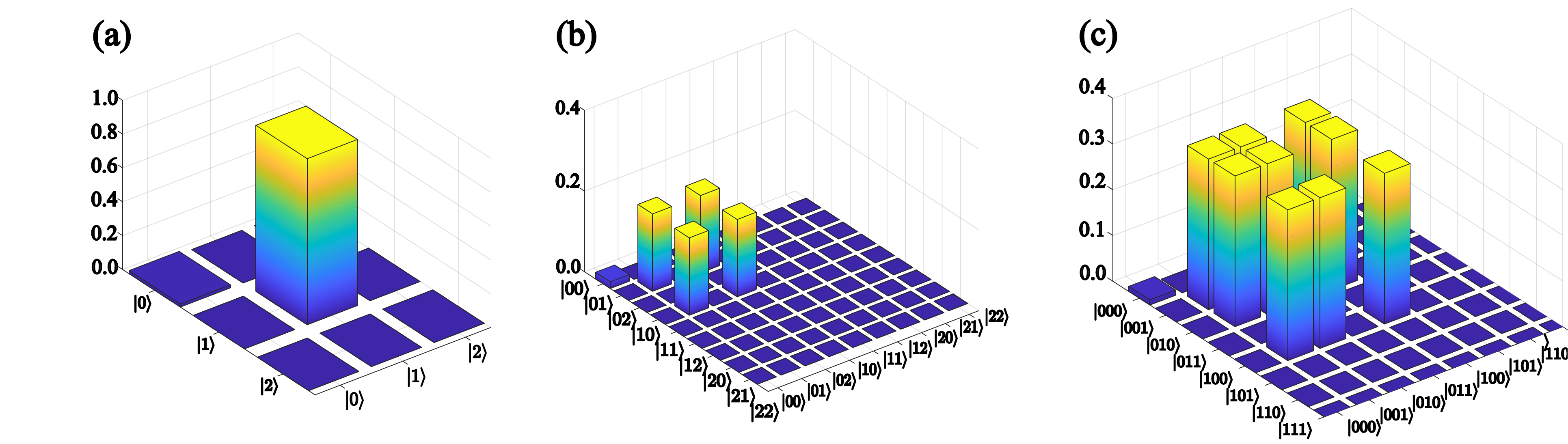}
\caption{Reconstructed density matrices associated with the modes $\omega_{1}$, i.e. $\rho_{\omega_{1}}$, for the case where the system is composed by $N=1$ (a), $N=2$ (b) and $N=3$ (c) multimode cavities. At the specific state generation times $t_S$, $t_B$, and $t_W$, the dominant amplitudes correspond to the states $\ket{1_{\omega_{1}}}$, $\ket{\Psi^{+}_{\omega_{1}}}$ and $\ket{W_{\omega_{1}}}$, respectively.}
\label{fig:fig3}
\end{figure*}

In Fig.~\ref{fig:fig2}, we show the numerical calculations of the above mentioned protocol. Here, we compute the population evolution of states $\ket{\Psi(0)}$, and states $\ket{\Psi}_S$, $\ket{\Psi}_B$, and $\ket{\Psi}_W$ given in Eqs.~(\ref{Eq9}). The parametric interaction can produce either uncorrelated photon states of different frequency or identical entangled states of modes belonging to distinct cavities. Furthermore, the simulations show that the state generation time decreases as $1/N$. This can be explained by analysing the structure of Eqs. (\ref{Eq8}). As the effective Hamiltonians describe a quantum dynamics in a reduced $2$-dimensional Hilbert space, the matrix elements between the initial state $\ket{\Psi(0)}$ and $\ket{\Psi}_S$, $\ket{\Psi}_B$, and $\ket{\Psi}_W$ are proportional to the normalization of the desired state, that is, $\sqrt{N}\times\sqrt{N}$, where $N=1$ stands for single photons, $N=2$ for Bell states, and $N\ge3$ for $W$ states. In other words, the matrix elements of the effective Hamiltonians are proportional to the number of multimode cavities. 
By considering the following parameters for the QRS, $\omega_{\rm{cav}}=2\pi\times13.12$~GHz \cite{PhysRevA.80.032109}, qubit frequency $\omega_{q}=0.8 \omega_{\rm{cav}}$, and light-matter coupling strength $g=0.6 \omega_{\rm{cav}}$, we can estimate $|\chi_{10}|=0.8188$ and $|\chi_{21}|=1.235$. In addition, we choose $\omega_{1}^{n}=0.25\nu_{20}$, $\omega_{2}^{n}=0.75\nu_{20}$, $J_{1}^{n}=0.0075\nu_{20}$, and $J_{2}^{n}=0.0053\nu_{20}$. In this case, the state generation times are about $t_{S}\approx25.10(8)~$[ns], $t_{B}\approx12.55(4)~$[ns], $t_{W}\approx8.369(4)~$[ns] for $N=3$, and $t_{W}\approx6.28~$[ns] for $N=4$, see Fig.~\ref{fig:fig2}. 
\begin{table}[b!]
\begin{ruledtabular}
\begin{tabular}{lccc}
&$N=1$&$N=2$ &$N=3$\\
\hline
$\mathcal{F}(\rho_{\omega_{1}},\rho_{\omega_{2}})$&0.9898  &0.9818  &0.9832  \\ 
\hline 
$\mathcal{F}_S$&0.9892  & - &-  \\ 
\hline
$\mathcal{F}_B$&-  & 0.9945 &-  \\ 
\hline
$\mathcal{F}_W$&-  & - &0.9904  \\ 
\end{tabular}
\end{ruledtabular}
\caption{\label{table1} Summarized Fidelity values between the states $\rho_{\omega_{\ell}}$ obtained through of the master equation (\ref{Eq17}) with the fictitious states $\rho_{\rm{probe}}$ and $\rho_{\rm{tensor}}$ for the case where the QRS is coupled to $n=\{1,2,3\}$ multimodes cavity.}
\end{table}
\section{Copies of density matrices}\label{copies}
In the above section, we have demonstrated that our system can generate identical copies of pure microwave photon states ($N=1,2,3$). Here, we demonstrate that even including loss mechanisms our protocol can still generate copies of density matrices with high fidelity. Since our proposal includes an ultrastrongly coupled light-matter system, the dissipative dynamics will be described by the master equation \cite{PhysRevA.84.043832} 
\begin{eqnarray}
\label{Eq10}\nonumber
\dot{\rho}(t)&=& i[\rho(t),\mathcal{H}]+\sum_{\ell=1}^{N}\kappa_{\ell}\mathcal{D}[b_{\ell}]\rho(t) +\sum_{\ell=1}^{N}\kappa_{\ell}\mathcal{D}[c_{\ell}]\rho(t) \\
&+&\sum_{\sigma,\sigma>\sigma'}(\Gamma_{\kappa}^{\sigma\sigma'}+\Gamma_{\gamma}^{\sigma\sigma'}+\Gamma_{\gamma_\phi}^{\sigma\sigma'})\mathcal{D}[\ketbra{\sigma,p}{\sigma',p'}]\rho(t).
\end{eqnarray}
Here, $\mathcal{H}$ is the Hamiltonian of Eq.~(\ref{Eq2}) and $\mathcal{D}[O]\rho=1/2(2O\rho O^\dag - \rho O^\dag O-O^\dag O\rho)$ is the Liouvillian operator. Furthermore, $\kappa_{\ell}^{n}$ stands for photon loss rate for each cavity mode. $\Gamma_{\kappa}^{\sigma\sigma'}$, $\Gamma_{\gamma}^{\sigma\sigma'}$ and $\Gamma_{\gamma_\phi}^{\sigma\sigma'}$ are the dressed decay rates
associated with the QRS, and they are defined as
$\Gamma_{\kappa}^{\sigma\sigma'} =\frac{\kappa}{\omega_{\rm cav}}\nu_{\sigma\sigma'}|{\rm{X}}_{\sigma\sigma'}|^2$, $\Gamma_{\gamma}^{\sigma\sigma'} = \frac{\gamma}{\omega_q}\nu_{\sigma\sigma'}|\sigma^{x}_{\sigma\sigma'}|^2$ and $\Gamma_{\gamma_\phi}^{\sigma\sigma'} =\frac{\gamma_\phi}{\omega_q}\nu_{\sigma\sigma'}|\sigma^{z}_{\sigma\sigma'}|^2$, where $\kappa$, $\gamma$ and $\gamma_\phi$ are the bare photon leakage, relaxation, and depolarizing noise rates, respectively.\par
To study the robustness of our protocol under loss mechanisms, first we will examine the generation of copies of density matrices for the cases of $N=1,2,3$ multimode cavities. As mentioned in the previous section, the whole system is initialized in the state $\ket{\Psi(0)}=\ket{0,+}\bigotimes_{\ell,\ell'}^{N}\ket{0_{\ell},0_{\ell'}}$. Then, we let the system to evolve under Eq.~(\ref{Eq10}) for three different times: $t_S=\pi/(2\mathcal{J}_{2}^{1})$, $t_B=\pi/(4\mathcal{J}_{2}^{1})$, and $t_W=\pi/(6\mathcal{J}_{2}^{1})$, for $N=1$, $N=2$, and $N=3$ multimode cavities, respectively. Once the corresponding density matrix $\rho(t)$ is obtained, we trace over the QRS and modes $\omega_2$ ($\omega_1$) to obtain the reduced density matrix $\rho_{\omega_1}$ ($\rho_{\omega_2}$) which contains only degrees of freedom associated with the mode $\omega_1$ ($\omega_2$) distributed on different multimode cavities. Table \ref{table1}, first row, shows the fidelity between both reduced density matrices $\mathcal{F}(\rho_{\omega_1},\rho_{\omega_2})=\rm{Tr}(\rho_{\omega_1}\rho_{\omega_2})$. These results allow us to conclude that both quantum states are identical up to $99\%$ fidelity
for a single cavity, and up to $98\%$ fidelity for two and three cavities. Table \ref{table1} also shows the fidelities of generating the states of Eqs.~(\ref{Eq9}), that is, $\mathcal{F}_S={\rm Tr}(\rho(t_S)\rho_S)$, $\mathcal{F}_B={\rm Tr}(\rho(t_B)\rho_S)$, and $\mathcal{F}_W={\rm Tr}(\rho(t_W)\rho_S)$, where $\rho(t)$ have been numerically calculated from Eq.~(\ref{Eq10}). In Fig. \ref{fig:fig3}, we plot the reconstructed density matrices for each case. The high fidelities of our protocol are mainly due to the fast state generation times as compared with the loss rates. Our numerical calculations has been carried out with realistic circuit QED parameters at temperature $\rm{T}=15~\rm{mK}$ \cite{Sci.Rep.6.26720}. For the QRS decay rates we consider values $\kappa=2\pi\times0.10~\rm{MHz}$, $\gamma=2\pi\times15~\rm{MHz}$ and $\gamma_\phi=2\pi\times7.69~\rm{MHz}$ 
and for the cavities $\kappa_\ell^n=\kappa$.
\section{Entanglement Swapping Between Distant superconducting qubits}
\label{swapping}
\begin{figure}[t!]
\centering
\includegraphics[width=1\linewidth]{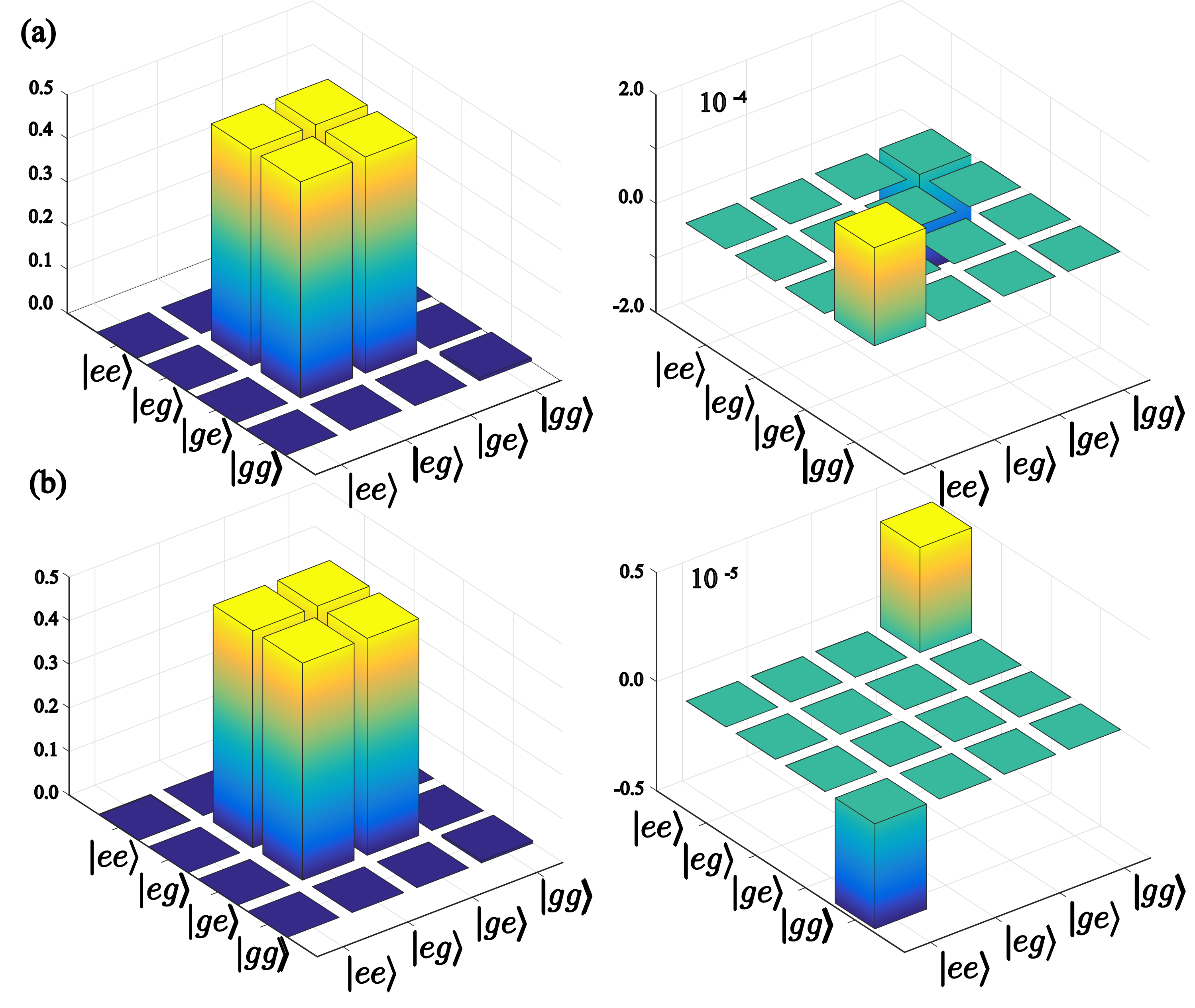}
\caption{Real and imaginary part of the reduced density matrix composed of the two qubits coupled to the field mode of frequency  $\omega_{1}$~(a) and mode $\omega_{2}$ (b). The fidelity between the simulated state with the Bell state $\ket{\Phi}=(\ket{eg} + \ket{eg)}/\sqrt{2}$ is (a) $\mathcal{F}=0.9960$ and (b)~$\mathcal{F}=0.9976$.}
\label{fig:fig4}
\end{figure}
In this section, we study the transfer of entanglement generated into the field modes towards distant superconducting circuits. Let us consider a pair of two-level systems coupled at the end of each cavity. As we shall see later in Sec. \ref{implementation}, our physical implementation will consider $\lambda/4$ transmission line resonators, and superconducting flux qubits to guarantee strong coupling between them. In such a case, we describe the system with the following Hamiltonian
\begin{eqnarray}
\label{Eq17}
\mathcal{H}_{{\rm ES}} = \mathcal{H} + \sum_{\ell=1}^{2}\frac{\omega_{q\ell}^{n}}{2}\sigma^{z}_{\ell} + \sum_{\ell=1}^{2}\lambda_{\ell}\sigma^{x}_{\ell}(\mathcal{B}_{\ell}^{\dag} + \mathcal{B}_{\ell}),
\end{eqnarray}
where $\mathcal{H}$ is the Hamiltonian defined in Eq. (\ref{Eq2}). Moreover, $\sigma^{x}_{\ell}$ and $\sigma^{z}_{\ell}$ are Pauli matrices describing the two-level systems. $\mathcal{B}_{\ell}$ is the extended cavity operators which, depending on whether the superconducting qubits are coupled to the first or the second mode, can be $\mathcal{B}_{\ell}=b_{\ell}, c_{\ell}$, respectively. Finally, $\lambda_{\ell}$ is the coupling strength between the qubit and the field mode. The system dynamics is described by the following master equation
\begin{eqnarray}
\label{Eq18}\nonumber
\dot{\rho}(t) = [{\rm{Eq. (10)}}] +\sum_{\ell=1}^{N}\gamma_{\ell}\mathcal{D}[\sigma^{-}_{\ell}]\rho(t) + \sum_{\ell=1}^{N}\gamma_{\phi_\ell}\mathcal{D}[\sigma^{z}_{\ell}]\rho(t).\\
\end{eqnarray}
The last two terms describe the loss mechanisms acting on the two-level system, i.e., relaxation on the qubit at a rate $\gamma$ and depolarizing noise at rate $\gamma_{\phi}$. The entanglement swapping protocol is the following; we initialize the whole system in its ground state
\begin{eqnarray}
\label{Eq19}
\rho_{0} =\ketbra{0,+}{0,+}\bigotimes_{\ell,\ell'}^{N}\ketbra{0_{\ell},0_{\ell'}}{0_{\ell},0_{\ell'}}\bigotimes_{\ell}^{N}\ketbra{g_{\ell}}{g_{\ell}},
\end{eqnarray}
We dispersively couple the two-level systems with the field modes on the cavities ($\omega_{\ell}^{1,2}\gg\omega_{q\ell}$). Then, we drive the QRS to prepare in the second excited state $\ket{2,+}$ 
\begin{eqnarray}
\label{Eq20}
\rho_{1} =\ketbra{2,+}{2,+}\bigotimes_{\ell,\ell'}^{N}\ketbra{0_{\ell},0_{\ell'}}{0_{\ell},0_{\ell'}}\bigotimes_{\ell}^{N}\ketbra{g_{\ell}}{g_{\ell}},
\end{eqnarray}
This state is the initial condition of our scheme. Afterwards, we let the system evolve under the Hamiltonian in Eq. (\ref{Eq17}). Due to the dispersive qubit-resonator interaction, the two-level systems do not evolve. After a time $t=\pi/(2\mathcal{J}_{\rm{eff}})$, the density matrix of the system reads  
\begin{eqnarray}
\label{Eq21}
\rho_{2} =\ketbra{0,+}{0,+}\bigotimes_{\ell}^{N}\ketbra{\Psi^{+}_{\omega_{\ell}}}{\Psi^{+}_{\omega_{\ell}}}\bigotimes_{\ell}^{N}\ketbra{g_{\ell}}{g_{\ell}}.
\end{eqnarray}
The next step is to avoid the generated photons coming back to the QRS. To achieve it, we tune far-off resonance the QRS and the resonators by changing the qubit frequency that belongs to the QRS. Afterwards, we put into resonance the external two-level system with either $\omega_{\ell}^{1}$ or $\omega_{\ell}^{1}$ field modes. In such a case, for a time $t=\pi/(2\lambda_{\ell})$, the system evolves to 
\begin{eqnarray}
\label{Eq22}
\rho_{3}=\ketbra{0,+}{0,+}\bigotimes_{\ell=1}^{N}\ketbra{0_{\omega_{1}^{\ell}}}{0_{\omega_{1}^{\ell}}}\otimes\ketbra{\Psi_{\omega_{2}^{\ell}}}{\Psi_{\omega_{2}^{\ell}}}\otimes\ketbra{\Phi}{\Phi},\\
\rho_{3}=\ketbra{0,+}{0,+}\bigotimes_{\ell=1}^{N}\ketbra{0_{\omega_{1}^{\ell}}}{0_{\omega_{1}^{\ell}}}\otimes\ketbra{\Psi_{\omega_{2}^{\ell}}}{\Psi_{\omega_{2}^{\ell}}}\otimes\ketbra{\Phi}{\Phi}
\end{eqnarray}
Here, $\ket{\Phi}=(\ket{g_{1}e_{2}} + \ket{e_{1}g_{2}})/\sqrt{2}$ is a Bell state of the pair of qubits. Fig. \ref{fig:fig4} shows the real and imaginary part of the reduced density matrix for the pair of qubits after performing the protocol. As the figure shows, even though the loss mechanisms act on the system, the entanglement of the modes can be transferred to the qubits with high fidelity. For the two-level systems coupled to the first mode ($\omega_{1}^{\ell}$), the fidelity is $\mathcal{F}=0.9960$, and $\mathcal{F}=0.9976$ when the qubit is resonant with the second mode ($\omega_{2}^{\ell}$). This transfer occurs at the time scale of  $t_{{S}_{1}}=23.08~$[ns] and $t_{{S}_{2}}=16.32~$[ns], respectively.
\section{Implementation in circuit QED}
\label{implementation}
\begin{figure}[t!]
\centering
\includegraphics[width=1\linewidth]{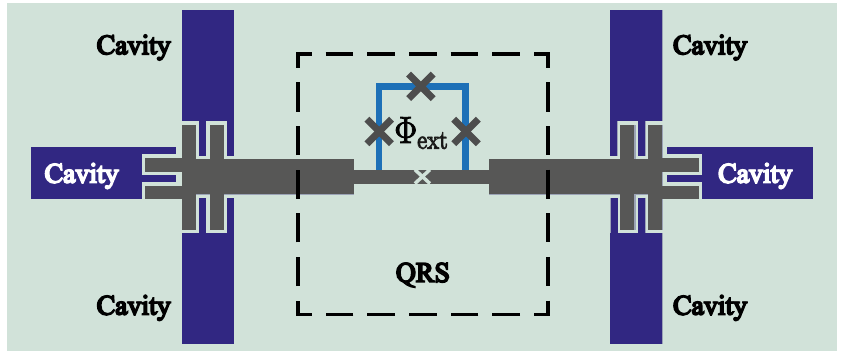}
\caption{(Color online). Schematic illustration of our superconducting circuit implementation. Here, the QRS is composed of a $\lambda/2$ transmission line resonator interacting with a superconducting flux qubit located at the middle point to achieve the USC regime. In addition, the $\lambda/2$ resonator is coupled at its edges to multimode transmission lines via capacitors.}
\label{fig:fig5}
\end{figure}
We depict the schematic implementation of our system in Fig.~\ref{fig:fig5}. The circuit is composed of a $\lambda/2$ transmission line resonator (TLR) galvanically coupled to a four-junction flux qubit at the middle of the resonator. Moreover, at the edges of this $\lambda/2$ resonator, one may couple two (up to six) additional $\lambda/4$ TLR via capacitances. The capacitive coupling follows the same procedure as in Ref.~\cite{1}. In such a case, the finger pattern at the end of these resonators form the capacitive coupling. The orthogonal arrangement between the multimodes cavity reduces the crosstalk between these resonators, reducing the cavity-cavity interaction. The Lagrangian representing this situation reads
\begin{eqnarray}
\label{circuit_lagrangian}
\mathcal{L} =\mathcal{L}_{\rm{QRS}} + \mathcal{L}_{\rm{c}} +  \mathcal{L}_{I},
\end{eqnarray}
where, $\mathcal{L}_{\rm{QRS}}$ is the QRS Lagrangian constituted by the $\lambda/2$ TLR coupled to a four-junction flux qubit, $\mathcal{L}_{\rm{c}}$ is the multimode $\lambda/4$ transmission line resonator Lagrangian, whereas $\mathcal{L}_{I}$ stands for the resonator-resonator coupling Lagrangian. The QRS Lagrangian is given by
\begin{eqnarray}
\label{QRS_lagrangian}\nonumber
\mathcal{L}_{\rm{QRS}} &=& \int_{0}^{d}dz\bigg[\frac{c}{2}[\partial_{t}\psi(z,t)]^2 - \frac{1}{2l}[\partial_{z}\psi(z,t)]^2\bigg]\\
&+&\sum_{k=1}^{4}\bigg[\frac{C_{J,k}}{2}\dot{\varphi}_{k}^2+E_{J,k}\cos\bigg(\frac{\varphi_{k}}{\phi_{0}}\bigg)\bigg].
\end{eqnarray}
Here, $c$ and $l$ are the capacitance and inductance per unit of length of the resonator, while $C_{J,k}$ and $E_{J,k}$ are the capacitance and energy describing the $k$-th Josephson junction. The multimode resonator Lagrangian is given by 
\begin{eqnarray}
\label{resonator_lagrangian}\nonumber
\mathcal{L}_{\rm{c}} &=& \sum_{\ell=1}^{N}\int_{0}^{d}dz\bigg[\frac{c_{\ell}}{2}[\partial_{t}\phi_{\ell}(z,t)]^2 - \frac{1}{2l_{\ell}}[\partial_{z}\phi_{\ell}(z,t)]^2\bigg]\\
&+&\frac{C_{r}}{2}[\partial_{t}\phi_{\ell}(d,t)]^{2} + \frac{C_{r}}{2}[\partial_{t}\phi_{\ell}(0,t)]^{2},
\end{eqnarray}
Finally, $\mathcal{L}_{I}$ is the interaction Lagrangian given by
\begin{eqnarray}
\label{interaction_lagrangian}
\mathcal{L}_{I} = -C_{r}\bigg[\dot{\phi}_{1}(d,t)\dot{\psi}(0,t) + \dot{\psi}(d,t)\dot{\phi}_{2}(0,t)\bigg].
\end{eqnarray}
\subsection{Rabi system Hamiltonian}
For this derivation we assume $E_{J,1}=E_{J,2}=E_{J}$, $E_{J,3}=\alpha E_{J}$ and $E_{J,4}=\gamma E_{J}$. Moreover, the fluxoid quantization relation on the superconducting loop is given by
\begin{eqnarray}
\label{fluxoid_relation}
\varphi_{1}-\varphi_{2} + \varphi_{3} + \varphi_{4}=-2\pi f_{x},
\end{eqnarray}
here, $f_{x}$ is the frustration parameter defined as $f_x=\phi_{\rm{ext}}/\Phi_{0}$. On the other hand, as the fourth junction is small enough in comparison with the loop forming the flux qubit, the superconducting phase difference along this junction corresponds to the phase difference of the $\lambda/2$ TLR, i.e. $\varphi_4=\Delta\psi$ \cite{PhysRevLett.108.120501}. Thus, the Lagrangian takes the following form
\begin{eqnarray}
\label{QRS_lagrangian2}\nonumber
\mathcal{L}_{\rm{QRS}} &=& \int_{0}^{d}dz\bigg[\frac{c}{2}[\partial_{t}\psi(z,t)]^2 - \frac{1}{2l}[\partial_{z}\psi(z,t)]^2\bigg]\\\nonumber
&+&\frac{C_{J}}{2}\bigg[\dot{\varphi_{1}}^2+\dot{\varphi}_{2}^2+\alpha(\dot{\varphi_{2}}-\dot{\varphi_{1}}-\Delta\dot{\psi})^2 + \gamma\Delta\dot{\psi}^2\bigg]\\\nonumber
&+&E_J\bigg[\cos\bigg(\frac{\varphi_{1}}{\phi_{0}}\bigg) + \cos\bigg(\frac{\varphi_{2}}{\phi_{0}}\bigg) + \gamma\cos\bigg(\frac{\Delta \psi}{\phi_{0}}\bigg) \\
&+& \alpha\cos\bigg(\frac{\varphi_{2}-\varphi_{1} + \phi_{\rm{ext}}-\Delta \psi}{\phi_{0}}\bigg)\bigg].
\end{eqnarray}
We are assuming the superconducting phase on the loop is well localized, thus the potential energy can be expanded in powers of $\Delta\psi/\phi_{0}$ \cite{PhysRevLett.108.120501}, allowing us to express the QRS Lagrangian in the following form
\begin{eqnarray}
\label{QRS_lagrangian3}
\mathcal{L}_{\rm{QRS}}=\mathcal{L}_{\rm{r}} + \mathcal{L}_{\rm{q}} + \mathcal{L}_{\rm{qr}}
\end{eqnarray}
where $\mathcal{L}_{\rm{r}}$ is the Lagrangian of the resonator with an embedded junction
\begin{eqnarray}
\label{QRS_resonator}\nonumber
\mathcal{L}_{\rm{r}}&=&\int_{0}^{d}dz\bigg[\frac{c}{2}[\partial_{t}\psi(z,t)]^2 - \frac{1}{2l}[\partial_{z}\psi(z,t)]^2\bigg]\\
&+&\frac{C_{J}(\alpha + \gamma)}{2}\Delta\dot{\psi} + \gamma E_{J}\cos\bigg(\frac{\Delta \psi}{\phi_{0}}\bigg).
\end{eqnarray}
Moreover, $\mathcal{L}_{\rm{q}}$ is the usual three-junction flux qubit Lagrangian~\cite{Phys.Rev.B.60.15398}

\begin{eqnarray}
\label{QRS_qubit}
\mathcal{L}_{\rm{q}}&=& \frac{C_{J}}{2}\bigg[(1+\alpha)(\dot{\varphi_{1}}^2+\dot{\varphi}_{2}^2)-2\alpha\dot{\varphi_{2}}\dot{\varphi_{1}} \bigg]\\\nonumber
&+&E_J\bigg[\cos\bigg(\frac{\varphi_{1}}{\phi_{0}}\bigg) + \cos\bigg(\frac{\varphi_{2}}{\phi_{0}}\bigg)+ \alpha\cos\bigg(\frac{\varphi_{2}-\varphi_{1}+\phi_{\rm{ext}}}{\phi_{0}}\bigg)\bigg].
\end{eqnarray}
Finally, $\mathcal{L}_{\rm{qr}}$ is the qubit-resonator Lagrangian; this term has two contributions: capacitive and galvanic coupling, and reads
\begin{eqnarray}
\label{QRS_int1}\nonumber
\mathcal{L}_{\rm{qr}}=-\alpha C_{J}(\dot{\varphi_{1}}+\dot{\varphi_{2}})\Delta\dot{\psi} - \frac{\alpha E_{J}}{\phi_{0}}\sin\bigg(\frac{\varphi_{1}-\varphi_{2}+\phi_{\rm{ext}}}{\phi_{0}}\bigg)\Delta\psi.\\
\end{eqnarray}
In the flux qubit, the capacitive energy is smaller than the inductive energy \cite{PhysRevA.80.032109}. Thus, we neglect the capacitive term, obtaining
\begin{eqnarray}
\label{QRS_int2}
\mathcal{L}_{\rm{qr}}= -\frac{\alpha E_{J}}{\phi_{0}}\sin\bigg(\frac{\varphi_{1}-\varphi_{2}+\phi_{x}}{\phi_{0}}\bigg)\Delta\psi.
\end{eqnarray}
We obtain the Lagrangian for the transmission line resonator by computing its equation of motion. In such a case, the flux $\psi(z,t)$ obeys the wave equation whose solution for the $\lambda/2$ TLR is given by 
\begin{eqnarray}
\label{QRS_resonator1}
&&\psi(z,t) = \sum_{m}\mathcal{U}_{m}(z)\mathcal{G}_{m}(t),\\
&&\psi(z,t) = \sum_{m}\bigg[A_{m}\mathcal{C}_{m}(z-d/2) + B_{m}\mathcal{S}_{m}(z+d/2)\bigg]\mathcal{G}_{m}(t),
\end{eqnarray} 
where $k_{m}$ is the wave vector of the resonator with the embedded junction, which is obtained through the dispersion relation
\begin{eqnarray}
\label{QRS_resonator2}
k_{m}\tan\bigg(\frac{k_{m}d}{2}\bigg)=\frac{2l}{L_{J}}\bigg[1-\bigg(\frac{vk_{m}}{\omega_{p}}\bigg)^{2}\bigg],
\end{eqnarray}
with $v=\sqrt{\frac{1}{lc}}$ is the TLR wave velocity, $L_{J}=\gamma\phi_{0}^{2}/E_{J}$ is the Josephson inductance. Besides, $\omega_{p}=1/\sqrt{L_{J}C_{J}}$ is the plasma frequency of the embedded junction. Replacing the flux $\psi(z,t)$ on the Lagrangian given in Eq. (\ref{resonator_lagrangian}) we arrive at
\begin{eqnarray}
\label{QRS_resonator3}
\mathcal{L}_{\rm{r}}=\sum_{m}\bigg[\frac{\eta_{m}\dot{\mathcal{G}_{m}}(t)^2}{2} - \frac{\eta_{m}^{2}\omega_{m}^{2}\mathcal{G}_{m}^{2}(t)}{2}\bigg].
\end{eqnarray}
where $\eta_{m}$ is the effective capacitance \cite{New.Jour.Phys.14.075024}. By applying the Legendre transformation, we arrive at the classical Hamiltonian
\begin{eqnarray}
\label{QRS_resonator4}
\mathcal{H}_{\rm{r}} = \sum_{m}\bigg[\frac{\Pi_{m}^{2}}{2\eta_{m}} + \frac{\eta_{m}^{2}\omega_{m}^{2}G_{m}^{2}}{2}\bigg].
\end{eqnarray}
Here, $\Pi_{m}=\partial \mathcal{L}/\partial[\dot{G}_{m}]$ is the canonical conjugate momenta. We proceed to quantize the Hamiltonian promoting the following operators
\begin{eqnarray}
\label{QRS_resonator4}
\Pi_m=\sqrt{\frac{\hbar}{2\eta_m\omega_m}}(a_m^\dag+a_m),\\
G_m=i\sqrt{\frac{\hbar\eta_m\omega_m}{2}}(a_m^\dag-a_m).
\end{eqnarray}
Replacing these operators in the Hamiltonian $\mathcal{H}_{r}$ we arrive at the TLR quantum Hamiltonian
\begin{eqnarray}
\label{QRS_resonator5}
\mathcal{H}_{\rm{r}} = \sum_{m}\hbar \omega_{m}\bigg(a_{m}^{\dag}a_{m}+\frac{1}{2}\bigg).
\end{eqnarray}
Now, let us consider the Lagrangian of the four-junction flux qubit given in Eq. (\ref{QRS_qubit}). Near from the degeneracy point $ \phi_{x} = \phi_{0}/2$, the system can be truncated to the two lowest eigenstates, whose Hamiltonian is given by
\begin{eqnarray}
\label{QRS_qubit1}\mathcal{H}_{\rm{q}}=\frac{\hbar \omega_{\rm{q}}}{2}\sigma^{z}
\end{eqnarray}
where, $\omega_{\rm{q}}=\sqrt{\Delta^{2}+\varepsilon^{2}}$, with $\Delta$ the qubit gap, and $\varepsilon=2I_{p}(\phi_{x}-\phi_{0}/2)$, where $I_{p}$ is the persistent current on the superconducting loop. Furthermore, the interacting Lagrangian given in Eq. (\ref{QRS_int2}) can be written in the two-level basis, in such case, the quantized Hamiltonian reads
\begin{eqnarray}
\label{QRS_int3}
\mathcal{H}_{\rm{qr}}= i\frac{\alpha E_{J}\Delta \mathcal{U}_{m}}{\phi_{0}}\sqrt{\frac{\hbar\eta_m\omega_m}{2}}\sigma^{x}(a_{m}-a_{m}^{\dag})
\end{eqnarray}
Thus, by considering the first mode of the resonator, we obtain the QRS Hamiltonian as follows
\begin{eqnarray}
\mathcal{H}_{\rm{QRS}}= \hbar\omega_{\rm{cav}}a^{\dag}a + \frac{\hbar\omega_{q}}{2}\sigma^{z} +  \hbar g\sigma^{x}(a^{\dag} + a).
\end{eqnarray}
\subsection{Multimode cavity Hamiltonian}
To obtain the Hamiltonian of the multimode cavities, let us consider the Lagrangian given in Eq. (\ref{resonator_lagrangian}) for $N=2$ resonators
\begin{eqnarray}
\label{resonator_lagrangian2}\nonumber
\mathcal{L}_{\rm{c}} &=& \sum_{\ell=1}^{N}\int_{0}^{d}dz\bigg[\frac{c_{\ell}}{2}[\partial_{t}\phi_{\ell}(z,t)]^2 - \frac{1}{2l_{\ell}}[\partial_{z}\phi_{\ell}(z,t)]^2\bigg]\\
&+&\frac{C_{r}}{2}[\partial_{t}\phi_{1}(d,t)]^{2} + \frac{C_{r}}{2}[\partial_{t}\phi_{2}(0,t)]^{2},
\end{eqnarray}
For the specific implementation, we consider boundary conditions defining a $\lambda/4$ resonator which are given by
\begin{eqnarray}
\label{BCs}
-\partial_{z}\phi_{1}(0,t) = -\partial_{z}\phi_{2}(d,t)=0,\\
\partial_{t}\phi_{1}(z,t) = \partial_{z}\phi_{2}(z,t)=0.
\end{eqnarray}
In this case, the dispersion relation reads
\begin{eqnarray}
\label{resonator_mode}
q_{n,\ell}=\frac{1}{v_{\ell}l_{\ell}C_{r}}\cot(q_{n}d).
\end{eqnarray}
Finally, the quantum Hamiltonian for the multimode resonators is given by
\begin{eqnarray}
\label{resonator_hamiltonian2}
\mathcal{H}_{\rm{c}}=\sum_{\ell=1}^{N}\bigg[\hbar \omega_{1}^{\ell}\bigg(b_{\ell}^{\dag}b_{\ell}+\frac{1}{2}\bigg) + \hbar \omega_{2}^{\ell}\bigg(c_{\ell}^{\dag}c_{\ell}+\frac{1}{2}\bigg)\bigg].
\end{eqnarray}

\subsection{Driving the superconducting qubit}
We can drive the two-level system by applying a time-dependent magnetic field on the superconducting loop, see Fig.~\ref{fig:fig5}. In such case, the energy gap $\omega_{q}$ can be expressed as 
\begin{eqnarray}
\omega_{q}(t)=\sqrt{\Delta^{2}+\varepsilon^{2}(t)}
\end{eqnarray}
where, $\varepsilon(t)=\varepsilon_{\rm{DC}} + \varepsilon_{\rm{AC}}\cos(\omega_{L}t)$ is the time-dependent energy on the system, which contains DC and AC contributions \cite{Phys.Rev.A.75.063414}. For $\varepsilon_{DC}\gg\varepsilon_{AC}$, we can write the flux-qubit energy as 
\begin{eqnarray}
\omega_{q}=\sqrt{\Delta^{2}+\varepsilon_{DC}^{2}} +\frac{\varepsilon_{\rm{DC}}\varepsilon_{\rm{AC}}}{\sqrt{\Delta^{2}+\varepsilon_{DC}^{2}}}\cos(\omega_{L}t).
\end{eqnarray}
Thus, the flux-qubit driving Hamiltonian is given by 
\begin{eqnarray}
\mathcal{H}_{q}(t)=\frac{\omega_{q}}{2}\sigma^{z} +\Omega\cos(\omega_{L}t)~\sigma^{z}.
\end{eqnarray}

\section{Discussion and conclusion}\label{conclusion}
In summary, we have shown the usefulness of the QRS to generate photons under suitable configuration. Based on the selection rules and the anharmonicity present in the QRS, it is possible to find the specific matching condition for producing two-photon processes, analogous to the observed in the parametric frequency conversion. This condition allows us to generate in a deterministic manner uncorrelated or correlated photon states, Bell and W states. The protocol mentioned above, together with available optical to microwave photon converter technologies, may be a useful resource to perform tasks as distributed quantum computing or quantum cryptography.

On the other hand, the proposed protocol could work as a quantum random number generator (QRNG) in the microwave regime. Unlike the optical regime where QRNGs are based on single mode and polarization states of photons, our proposal considers multimode states of photons. As a consequence, we observe a quadratic increase in the amount of possible quantum random numbers that would be generated in comparison with the single-mode case. Moreover, due to the fact that our system generates simultaneously identical maximally entangled photonic states of different frequency, this state resembles a $N^{2}$-side dice, where each side is associated with the probability to find the photons of frequency $\omega_ {1} $ and $\omega_ {2} $ in one of the two modes in $N$ cavities. Thus, the multiphoton process mediated by a QRS occurring on multimode cavities provides an efficient way to produce quantum random numbers. This efficiency relies on two main aspects of our protocol. The former concerns with the collective effect producing a decrease of the generation time as the number of cavities increases, permitting to avoid the bias produced by the interaction of the system with the environment. The latter concerns the multimode configuration of our scheme. As we previously mentioned, the inclusion of the multimode systems allows us to increase the amount of possible quantum random numbers as the number of devices required decrease. Finally, we have also proposed a possible experimental implementation of our scheme considering near-term technology on circuit quantum electrodynamics in the ultrastrong coupling regime.
\section*{Acknowledgements} 
We thank Leong-Chuan Kwek for fruitful discussions. The authors acknowledge support from CEDENNA. F.A.C.-L., Financiamiento Basal para Centros Cient\'ificos y Tecnol\'ogicos de Excelencia FB.0807, Direcci{\'o}n de Postgrado USACH, FONDECYT grants No. 1150653 and No. 1140194, Spanish MINECO/FEDER FIS2015-69983-P, Basque Government IT986-16, and  Ram\'on y Cajal Grant RYC-2012-11391.

\end{document}